# Experimental demonstration of photonic quantum ratchet

Chi Zhang,[1] Chuan-Feng Li[*,1] and Guang-Can Guo[1]

[1]*Key Laboratory of Quantum Information,*

*University of Science and Technology of China, CAS, Hefei, 230026, China*

(Dated: September 10, 2012)

## Abstract

We created a potential for light with a phase mirror and then experimentally realized a photonic quantum ratchet in an all-optical system, in which ratchet effects can be observed with the naked eye up to more than 22 steps, and quantum resonance can be demonstrated. Our method also provides a new means to simulate quantum particles with classical light, and it can be applied to investigate many other quantum phenomena.

---

[*] email:cfli@ustc.edu.cn

Ratchet [1, 2] is a spatially periodic system in which the directed motion of particles can be generated without a bias force. The system can be applied to various delicate devices, such as new electron pumps [3], biological motors [4, 5], magnetic flux manipulator in artificial nanostructured superconductors [6–8], etc. The quantum ratchet [9–12], which is a quantum extension of this model, has triggered a considerable amount of interest, motivated by the exploration of basic quantum phenomena. Many types of quantum ratchets are based on the quantum delta-kicked rotor [15], which is a paradigm of quantum chaos [16] and plays a crucial role in many related studies, such as dynamical localization [17] and quantum resonance (QR) [18, 19]. Here we created a delta-kicked photonic quantum ratchet accelerator in an all-optical system, with a slowly varying grating like phase mirror as a potential for light. Remarkably, directed motion and quantum resonance can be observed with the naked eye for more than 22 steps. Our method can also be used to simulate quantum particles [20, 21] and optical blackholes [22, 23] with classical light, and has potentially applications that solving the evolution of wave functions like a quantum computer [24].

In the early studies of ratchet effects, stochastic fluctuations [14, 25] played an important role in generating the directed motion. However, recent studies had realized Hamiltonian quantum ratchets [26], where the transportation was driven by Hamiltonian chaos [16], in the absence of noise. Assume that there is a one-dimensional particle in a ratchet potential that is described by the Schroedinger equation [11, 15]

$$i\tilde{\hbar}\frac{\partial \psi}{\partial t} = -\frac{\tilde{\hbar}^2}{2}\frac{\partial^2 \psi}{\partial x^2} + Kv(x)\sum_{n=0}^{\infty}\delta(t-n)\psi, \qquad (1)$$

where x is the spatial coordinate, K is the effective potential strength, the sawtooth-shaped ratchet potential *v(x) = sin x + α sin(2x + φ)* is periodically flashing with delta kicks and n represents the number of kicks. To set the spatial period of the potential and the temporal period of the flashing unity, the effective Planck constant of the system is assumed to be

$$\tilde{\hbar} = 8\omega_R T, \qquad (2)$$

where *T* is the flashing period and $\omega_R = \hbar k_l^2/2m$ is the recoil frequency of the system, with *m* the mass of the particle and *k* the wave number giving a potential period *(2kL)$^{-1}$*.

The directed current can be established since the desymmetrization of Floquet eigenstates and the violation of all the relevant symmetries [27, 28]. The symmetry of the system is related to the parameter *φ*, when *φ =0, π,* there is no symmetry breaking, the normal quantum behavior in this system is dynamic localization, in which the transfer of energy to the system ceases after the quantum break time. However, for some certain flashing periods, QR occurs, and the quantum ratchet accelerator [29] can be generated, the kinetic energy increases quadratically with time. It seems as though the potential can always kick the particle at the right time to accelerate the system in the most rapid way [31]. It is similar to classical resonance, which occurs when the frequency of the driving force is equal to the eigen frequency of the system. A QR occurs when the flashing frequency is commensurate with the recoil frequency; more precisely, it is determined by the effective Planck constant [12]

$$\widetilde{\hbar} = 4\pi r/s, \tag{3}$$

where $r$ and $s$ are mutually prime integers. Cases with small $s$ and large $s$ are called low-order QR and high-order QR, respectively.

QRs and the ratchet effects have been experimentally studied in atomic systems [9, 18, 19, 31], and the potential was established by an optical lattice. However, direct observation of QR dynamics require an initial states with a long coherence width [12], which may be a strict condition for an atomic system. On the other hand, previous research on optical kicked system [30] indicates a possibility to study ratchet effects in optical system. Here we adopt a new scheme in an all-optical system to realize the photonic quantum ratchet and observe phenomena such as QR; the ratchet effects can be observed directly with the naked eye for 22 kicks so that the dynamics and mechanism are more easily understood.

Our experimental setup is described in Figure 1. A phase mirror served as the flashing ratchet potential and the particle is the photon of a laser beam reflected between a plane mirror and the phase mirror (see Methods for the description of that phase mirror as a delta-kicked potential for light). The beam is reflected between these two mirrors, so it will encounter the phase mirror again and again with a certain temporal interval. Therefore, if we only consider the motion of the beam inside the same plane with the mirrors, the phase mirror can be regarded as a periodically flashing potential. The potential strength, which is determined by the etched depth of the phase mirror, varies with x. Thus, the ratchet effect is in the momentum and kinetic energy in x direction, and the intensity distribution of the beam in the x axis is the probability distribution of the particle because their motions are described by functions in the same form. In addition, to observe the ratchet effect of each kick clearly, the y positions of these kicks should be separated from one another. This can be realized by introducing a small, nonzero initial incidence angle on the y direction to the beam (Fig. 1), this will not affect the motion in x direction while each kick can be identified by a different y coordinate. However, to observe the ratchet effect, we need to know the momentum rather than the position. To obtain the momentum, a cylindrical lens should be used to convert the pattern of the position distribution into that of the momentum distribution, because a lens can concentrate all the beams in the same direction, or momentum, to the same point on its focal plane. In addition, a cylindrical lens only concentrates beams in one direction but does not affect the other direction, so each step is separated as before. Finally, we integrated the plane mirror with the cylindrical lens; that is, we coated the flat side of the lens. Consequently, the beam could then be reflected between the phase mirror and the flat side of the cylindrical lens to generate the ratchet effect. Part of the light that passes though the coating side will be converted to momentum distribution on the focal plane of the lens, which can then be observed on a screen or recorded with a CCD camera.

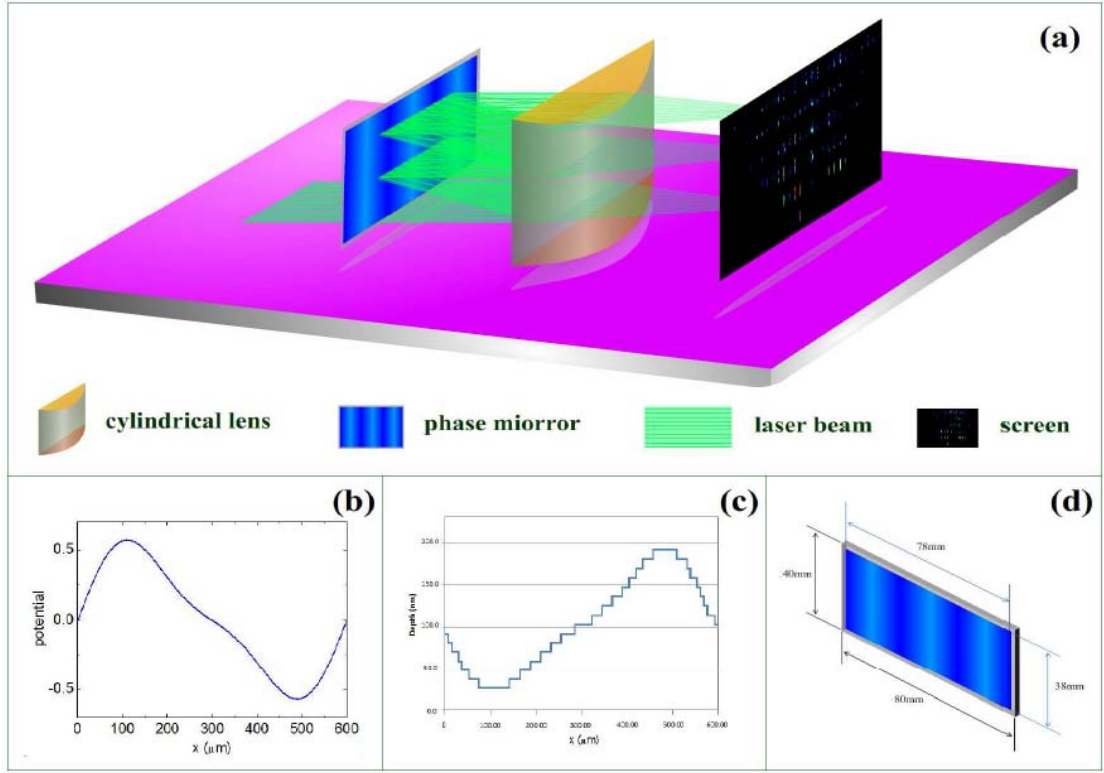

FIG. 1: (a). Experimental setup. The laser beam is shaped as an ellipsoid (approximately 6 mm × 1 mm) with two pairs of cylindrical lenses; it then enters the space between the phase mirror and the flat side of the cylindrical lens. Every time it encounters the cylindrical lens, approximately 95% is reflected and 5% passes through. Because the incident angle is not zero, the laser beam hits a different vertical position every time, which can identify the number of kicks on the phase mirror. In addition, when the transmitted 5% beam is concentrated on the focal plane of the cylindrical lens (focal length 300 mm), then we observe the momentum of each step directly on a screen there. (b). One spatial period of the ideal sawtooth-shaped potential of the phase mirror, $v(x) = \sin x + \alpha \sin 2x$, where $\alpha = 0.3$ and $x$ is normalized by a spatial period of 600 µm. (c). Etched depth of the surface of the phase mirror; the depth is approximated to 16 discrete levels. (d). Appearance of the phase mirror. The dimension is 80 mm × 40 mm, and there is a 1 mm width transparent belt on every edge; the laser beam is transmitted from the bottom belt. The phase mirror is made of fused silica, and coated with an all-reflecting film after being etched.

In our experiment, the potential was set to be $v(x) = \sin x + \alpha \sin 2x$, with $\alpha = 0.3$, the spatial period was 600 µm, and the effective potential strength K = 1. The operating wavelength was 532 nm, and the reflectivity of the coating on the flat side of the cylindrical lens was 95%. The temporal period was contr-

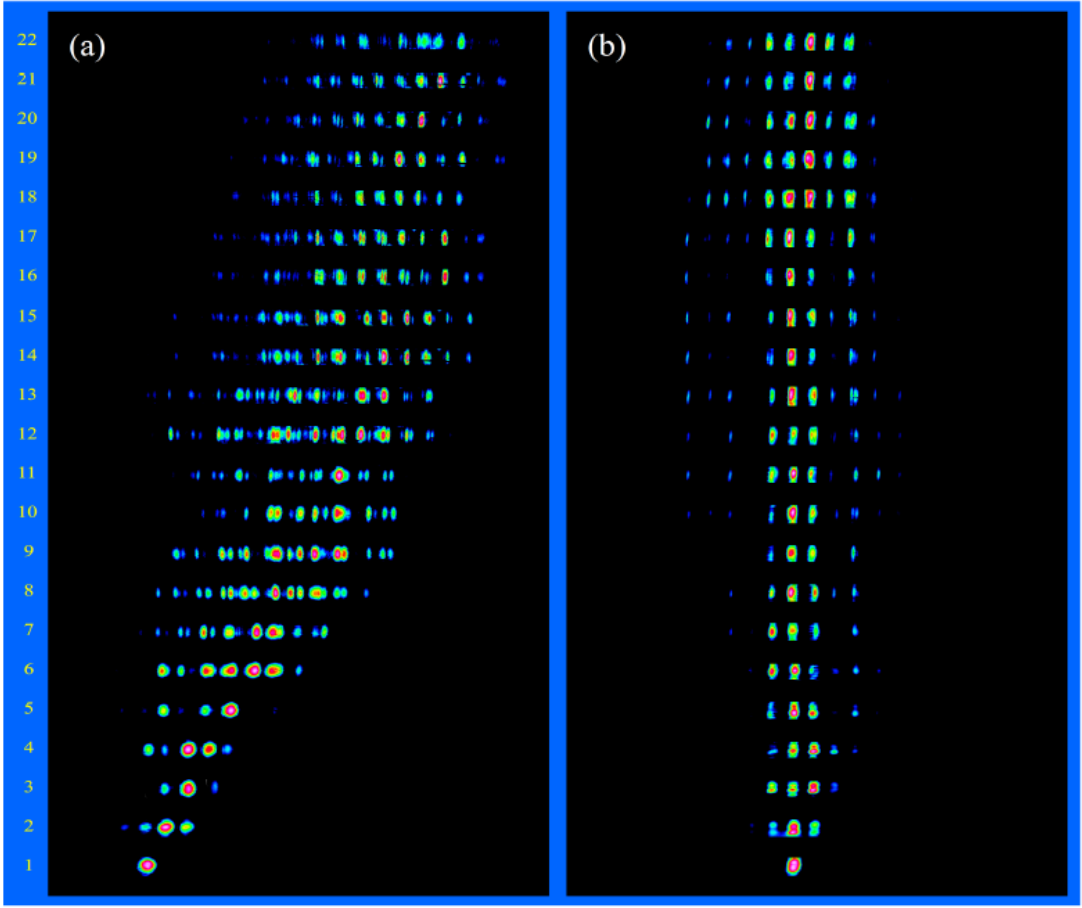

FIG. 2: A CCD image on the focal plane of the cylindrical lens. The background is black and the colored parts indicate the power of the pixels there. We found that the momentum can only be measured in discrete values because of the periodical potential. (a). The effective Planck constant $\tilde{\hbar} = 0.5\pi$, the mean momentum increases linear with kicks. (b). $\tilde{\hbar} = 0.35\pi$, the mean momentum does not increase obviously.

-olled by the distance between the phase mirror and the cylindrical lens. According to Eq. 2, the effective Planck constant $\tilde{\hbar}$ in this optical system is given by

$$\tilde{\hbar} = \frac{4\hbar k_l^2 T}{m} = 2\pi \frac{\lambda L}{l^2}, \qquad (4)$$

where $m = \lambda c$ is the mass of photon, $\lambda$ is the wavelength, $l$ is the spatial period of the phase mirror and $L$ is the distance between the phase mirror and the cylindrical lens. $L$ is the only variable in our experiment, so we can vary the distances to change the value of the effective Planck constant $\tilde{\hbar}$, and observe the ratchet effects and QR.

The incident beam was modulated into an elliptical shape with a wide x and narrow y distribution, and then entered the space between the phase mirror and cylindrical lens. First, we set $\tilde{\hbar} = 0.5\pi$ and $\tilde{\hbar} = 0.35\pi$, and took photographs of the pattern at several positions of the focal plane with a CCD camera, these photos were spliced together to obtain a complete image of the first 22 steps of the ratchet effect, which is shown in Fig. 2. It is obvious that the momentum can only be measured in discrete values because of the periodical potential, and it can also be seen that in case of $\tilde{\hbar} = 0.5\pi$, the photons were kicked to one side, whereas when $\tilde{\hbar} = 0.35\pi$, the photons were vibrating left and right.

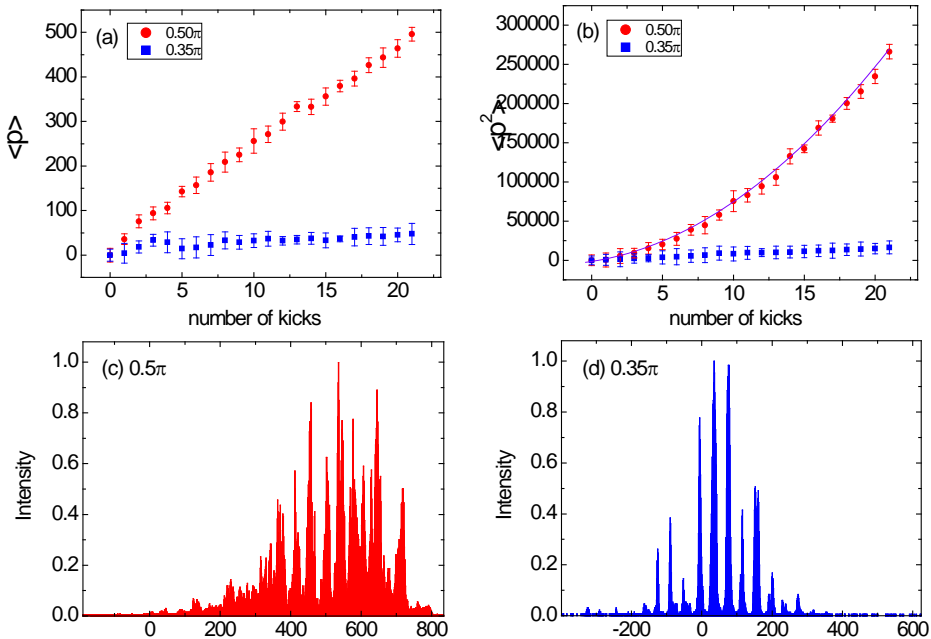

FIG. 3: (a). Mean momentum of the ratchet in the first 22 steps. The horizontal axis represents the number of kicks, and the vertical axis is the momentum, measured by pixel. $\tilde{\hbar} = 0.5\pi$ (red) shows a resonance while $\tilde{\hbar} = 0.35\pi$ (blue), shows random fluctuation. (b). The square momentum in the first 22 steps, $\tilde{\hbar} = 0.5\pi$ (red), it grows quadratically with number of kicks (the violet line is the quadratically fit of those points), and $\tilde{\hbar} = 0.35\pi$ (blue), it does not grow obviously. (c) and (d). The momentum distribution in the 22nd step when $\tilde{\hbar} = 0.5\pi$ and $\tilde{\hbar} = 0.35\pi$, respectively. The intensity has been normalized with the CCD camera program.

Next, the intensity on every pixel was recorded to calculate the directed motion more precisely. We compared the ratchet effects with two different $\tilde{\hbar} = 0.5\pi$ and $\tilde{\hbar} = 0.35\pi$, and the results are shown in Fig. 3. (a) reveals the mean momentum in every steps, and it increases rapidly when $\tilde{\hbar} = 0.5\pi$ while randomly changing when $\tilde{\hbar} = 0.35\pi$. (b) describes the increasing of the square momentum for these $\tilde{\hbar}$, when $\tilde{\hbar} = 0.5\pi$, it increases quadratically with number of kicks. (c) and (d) are the momentum distribution after 22 steps for $\tilde{\hbar} = 0.5\pi$ and $\tilde{\hbar} = 0.35\pi$, respectively.

Furthermore, we varied the distance $L$, or $\tilde{\hbar}$, and measured the mean momentum at the 22nd step. The results are shown in Fig. 4, resonances appear when $\tilde{\hbar} = 0.5\pi, \pi, 1.5\pi, 2\pi$. In other cases, the mean momentum did not obviously increase.

The principle of our experiment is highly similar to that of atomic quantum ratchet. For particles which are constrained in one-dimensional optical lattice, the ratchet effects are caused by the interference of the wave function. In our experiment, on the other hand, the laser beam has extremely long coherent time, if we only take the evolution in the transversal plane into account, and regard the longitudinal dimension as time, similar to particles' wave function, photons also can be represented by the electromagnetic wave, because both the light wave and the wave function of particles are subject to the same principles of interference and superposition, the behavior of photon is exactly the same with those particles in optical lattice potential. Therefore, we called it photonic quantum ratchet, differs from classical ratchets, which are driven by noise, photonic quantum ratchet is generated by quantum resonance or Hamiltonian chaos because of the coherent of light. The only difference between photonic and atomic quantum ratchets is that the wavelength of photon is far longer than that of particle, so the spatial period of our phase mirror is 600 μm while the spatial period of optical lattice is usually less than 1 μm [31]. This is also an advantage of photonic quantum ratchet, it makes the ratchet easier to control and measure, when moving the cylindrical lens forward and backward, we can even see the pattern on the screen changing.

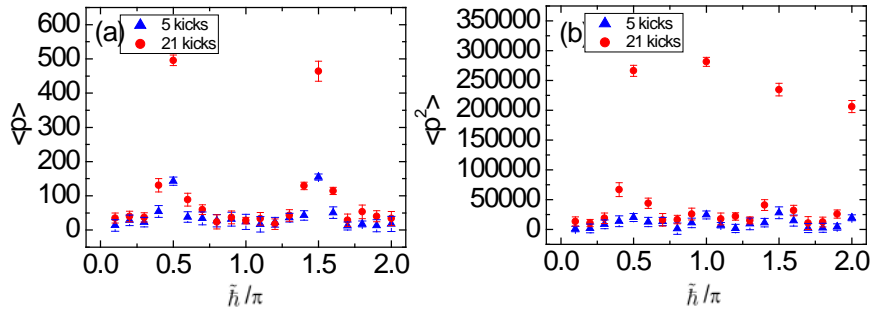

FIG. 4: Mean momentum after many kicks for different $\widetilde{\hbar}$, varying from 0 to $2\pi$. The red dots are experimental data after 21 kicks, and the blue dots are for 5 kicks. It can be obtained that when $\widetilde{\hbar} = 0.5\pi, \pi, 1.5\pi, 2\pi$, the systems are accelerated very rapidly, they are four resonance cases.

In addition, the resonance condition of $\widetilde{\hbar} = 4\pi r/s$ described by Eq. 4 is same with the condition of Lau effect [33, 34], more precisely, when two identical grating positioned one behind the other and backlighted by an extended white light source, colored fringes can be observed at infinity if the distance between two grating is $z = \frac{\alpha}{\beta} \frac{l^2}{2\lambda}$, where $\alpha$ and $\beta$ are integrals and $l$ is the grating period. They may have some connections, which suggest we try to extend Lau effect to atomic system.

In summary, we have realized a photonic quantum ratchet in an all-optical system and simulated quantum particles' behavior through classical light such that the probability distribution of the particle can be represented by the intensity of light. In addition, we have established a delta-kicked potential, phase mirror, for light. Its structure is similar to a grating, but the main difference is that the spatial period of a grating is usually in micrometers, commensurate with the wavelength of light. Therefore, if we use a grating as a ratchet potential, the effective Planck constant $\widetilde{\hbar}$ (Eq. 4) of the system will be commensurate with, which is extremely large so that even if $L$ varies several micrometers, $\widetilde{\hbar}$ will vary significantly and the evolution of the system will be totally unpredictable.

Our method provides an easy way to investigate quantum ratchets, and additional new schemes to improve the quantum ratchet could be studied further. For example, it is proposed that a pinch of disorder in the potential can likely improve the quantum ratchet [35] by reducing the dispersion rate of the ratchet transport. In other words, whether disorder-induced localization can slow down the diffusive spreading without affecting the mean momentum is an interesting problem that could be tested in our optical quantum ratchet. Moreover, as the light in phase mirror is similar to particles in potential, we can simulate one with the other, since light is more controllable and measurable, it will open up a new means of simulation.

**Methods**

**Phase mirror as a potential for light**

A phase mirror is a special mirror that can encode the phase of reflected light by etching the surface with a different depth. When light is reflected by the phase mirror, its phase in the wavefront will be encoded so that the propagation directions will change. If we only take the momentum inside the same plane with the phase mirror into account, it seems that the momentum changes after encountering the phase mirror. This is very similar to that particle's momentum change when encountering a potential, and the interaction time of the light reflection is infinitesimal; therefore, it can create an ideal delta-kicked potential for light.

Specifically, assume that a photon incident with an angle θ, the momentum inside the plane of phase mirror is $\frac{h}{\lambda}\sin\theta$, and if the depth in the area of that photon interacting with phase mirror is inhomogeneous, the reflection angle, or the momentum, will change, and $dp = \frac{h}{\lambda}d\sin\theta$. Suppose the phase shifts of two close points, $x_1$ and $x_2$, are $\phi(x_1)$ and $\phi(x_2)$, and the incidence and reflection angle are $\theta_1$ and $\theta_2$, respectively. We have

$$\phi(x_1) + \frac{dx\sin\theta_1}{\lambda} = \phi(x_2) + \frac{dx\sin\theta_2}{\lambda} \tag{5}$$

Therefore, $d\phi(x) + \frac{dx\, d\sin\theta}{\lambda}$. Compare to *dp*, we obtain

$$dp = h\frac{d\phi}{dx}. \tag{6}$$

On the other hand, for a particle in a delta-kicked potential V (x, t)= V (x)δ(t), the differential momentum is

$$dp = \int -(\frac{V(x)\delta(t)}{dx})dt = -\frac{dV}{dx}\int \delta(t)dt = -\frac{dV}{dx}. \tag{7}$$

Compared to the abovementioned equations, the phase shift can be regarded as a delta-kicked potential for the photon, $V(x) = -h\phi(x)$. This can not only be used for constructing a quantum ratchet, but it also provides a novel way to simulate quantum particles subject to delta-kicked potential by light.

According to a previous study on photonic crystal [21, 36], different dielectric, or index of refraction, act as different potential for photon (or light). The principle of the phase mirror is similar to that of photonic crystal, as they both change the phase on the wavefront. The difference is that a photonic crystal introduces a continuous potential, while a phase mirror introduces a delta-kicked potential.

**Acknowledgments** This research was supported by the National Basic Research Program (2011CB921200) and National Natural Science Foundation of China (Grant No. 60921091 and No. 11274289). C.Z. acknowledges the support by the Fund for Fostering Talents in Basic Science of the National Natural Science Foundation of China (No.J1103207).